# Inferring Informational Goals from Free-Text Queries: A Bayesian Approach

**David Heckerman and Eric Horvitz**
Microsoft Research
Redmond, WA 98052-6399
heckerma,horvitz@microsoft.com

## Abstract

People using consumer software applications typically do not use technical jargon when querying an online database of help topics. Rather, they attempt to communicate their goals with common words and phrases that describe software functionality in terms of structure and objects they understand. We describe a Bayesian approach to modeling the relationship between words in a user's query for assistance and the informational goals of the user. After reviewing the general method, we describe several extensions that center on integrating additional distinctions and structure about language usage and user goals into the Bayesian models.

## 1 Introduction

The management of uncertainty plays an important role in understanding the goals or intentions behind the words a person uses to communicate ideas. Problems with communication and understanding are exacerbated when people attempt to describe unfamiliar ideas and concepts. We present a Bayesian approach to interpreting queries composed by users of software applications searching for information about the means for achieving goals with the software. The problem of interpreting a user's informational goals given a query is typically one of reasoning under uncertainty. In the realm of assistance with consumer software applications, people typically are unfamiliar with the terms that expert user's or software designers may use to refer to structures displayed in a user interface, states of data structures, and classes of software functionality.

We take a Bayesian perspective on information retrieval (IR) for inferring the goals and needs of software users. The approach centers on the construction of probabilistic knowledge bases for interpreting user queries. Within the knowledge bases, variables representing a broad set of potential information goals influence the likelihood of a user generating different words in their queries for assistance. Our work comes in the context of the broader history of investigation of probabilistic and utility-theoretic methods in IR (Maron & Kuhns, 1960; Robertson, 1977; Cooper & Maron, 1978; Turtle & Croft, 1996; Keim, Lewis, & Madigan, 1997). Probabilistic analyses have been employed in a variety of ways in information retrieval, including work on ascribing a probabilistic semantics to notions of relevance associated with ranking strategies and learning from data. There has been growing interest in employing probabilistic graphical models to retrieval problems (Turtle & Croft, 1990; Fung & Del Favero, 1995). Discussion has continued in the IR community on the challenges and pitfalls of probabilistic representations and analysis (Cooper, 1994).

In contrast to previous data-centric, statistical approaches to information retrieval, we present our investigation of a more resource-intensive approach that relies on the framing and assessment of knowledge bases by human experts. Such handcrafting of knowledge bases is infeasible for grappling with massive retrieval problems like accessing information from large, heterogeneous databases like the World Wide Web. However, we found that the approach is appropriate for building a powerful retrieval tool for use in the focused context of providing help for users of the Microsoft Office suite of applications. The Microsoft Office suite of word processing, spreadsheet analysis, database, messaging and scheduling, and presentation applications is relied upon by millions of users. We found that the value associated with the quality of retrieval derived from a custom-tailored Bayesian IR system could justify the significant expense incurred by extensive user modeling.

We shall describe a general approach to modeling the probabilistic relationships between a user's words and their informational goals. First, we describe a basic model and updating procedure. We then discuss our work to introduce default probabilities to ease the assessment burden as well as leveraging additional structure in language to enhance the performance of the system following. Finally, we discuss how the assessment and inference strategies were scaled up to handle user assistance for applications in the Microsoft Office suite of desktop applications.

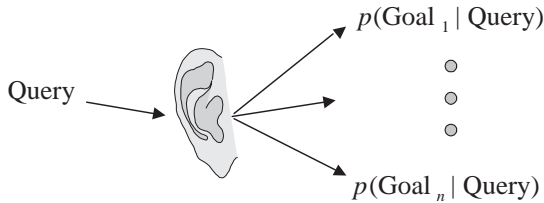

Figure 1: The overall goal of Bayesian assessment and inference is to infer a probability distribution over a set of concepts, given a query.

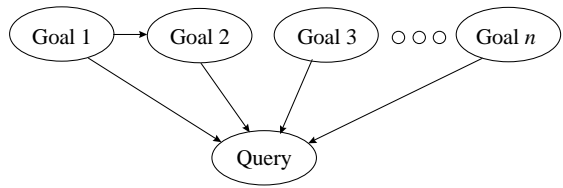

Figure 2: Bayesian network for diagnosing problems from a user query. This representation is intractable for knowledge acquisition given the large number of potential queries.

## 2 User goals, queries, terms

At the time we initiated our project in Bayesian information retrieval, managers in the Office division were finding that people were having difficulty finding assistance efficiently. Problems with accessing solutions to problems and for discovering new features were rooted in problems with terminology. As an example, users working with the Excel spreadsheet might have required assistance with formatting "a graph." Unfortunately, Excel had no knowledge about the common term, "graph," and only considered in its keyword indexing, the term "chart." One approach to this problem would have been to build sets of synonyms within the paradigm of keyword search. However, this approach would not have addressed the uncertain relationship between a user's query and informational goals. Consequently, we turned to a probabilistic approach for diagnosing a user's problems *given* a query.

In a general user-modeling approach to information retrieval, we seek to infer a probability distribution over concepts given a user's query or utterance. For example, if a user states "I need help with graphing," we would like to infer a probability distribution over a set of user goals. An appropriate probability distribution would likely be different for the query, "I want to change the way the graph looks." We pursued a Bayesian analysis of terms to generate a probability distribution over utterances.

Traditional approaches to Bayesian IR have employed word counting to gather probabilistic information on the relationships of words to concepts. With a user modeling approach, we have the ability leverage knowledge that is typically not available in text written for online assistance with software. We are interested in capturing the way people actually describe their problems.

In a general model we consider all possible queries and all possible goals, as captured by the Bayesian network in Figure 2. We structure this IR problem with a set of problem variables, representing a user's problems, influencing the probability distribution over plausible queries. As indicated in the figure, it may be important to represent dependencies among problems. In a high-fidelity model, we might seek to consider dependencies among problems over time and represent explicit temporal probabilistic relationships among problems. At run time, we observe the user's query and infer probabilities of concepts.

It is infeasible to represent the large number of distinct queries that might be composed by users. Instead, we developed an approximation. To simply the assessment and inference, we made the following assumptions:

- *Single user problem*: Only a single problem is active at any time.

- *Order irrelevance*: The order of terms in a query is disregarded.

- *Irrelevance of unrecognized terms*: Only words in the lexicon are included in the analysis.

Taken together, these assumptions define a *Bayesian term-spotting* methodology.

To ease the assessment burden we made a fourth assumption. We additionally assumed *term independence*: terms are independent of other terms, conditioned on the problem. A Bayesian network for the reformulated term-spotting problem is displayed in Figure 3. Clearly, assumptions of term independence and order irrelevance may lead to significant information loss, given the great importance of structure and dependencies among word in human communication. Nevertheless, we were interested in the quality of inference for Bayesian term spotting.

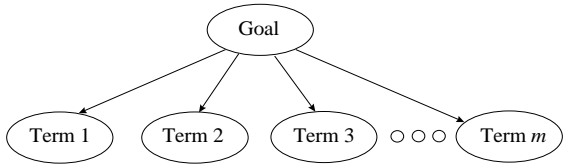

Figure 3: A reformulation of the problem into a Bayesian term-spotting analysis. We consider the presence or absence of words, discarding key information about the sequence of words and dependencies among words.

To construct the knowledge bases, we enlisted usability specialists to assist with the identification of the problems of users and their associated help topics, as well as terms that might be seen typically in queries conditioned on the problems of users. For a prototyping effort, we specified the set of problems associated with help topics for the Microsoft Excel application. For each problem, we identified sets of terms whose likelihood of appearing in a query would be influenced by the presence of a problem.

To simplify assessment, we employed a stemmer to reduce the number of terms by collapsing a multitude of derived forms into more basic roots or *lemmas*. For example, derived forms of the reference to *printing a document*, including "print," "printed" and "printing" were reduced to the root, "print."

Beyond consideration of root forms of words, special phrases and distinctions were added to consideration when deemed important. Such special distinctions included term with distinguished patterns of capitalization (e.g., "Word," as in Microsoft Word, and "word" are treated as distinct terms).

After constructing the problem and term distinctions, we pursued knowledge from experts on the probabilities of terms being used in a query conditioned on the existence of each problem. Specifically, we sought, for each term $t$, the conditional probability that the term would appear in a query given a user assistance goal $g$ linked to the terms, $p(t|g,\xi)$, where $\xi$ refers to the background state of knowledge of the person assessing the probability. To ease the assessment, a log scale was employed which discretized the conditional prob-abilities into a set of buckets which were separated by equivalent likelihood ratios.

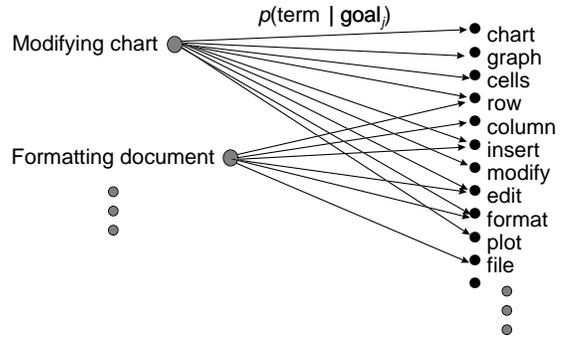

Figure 4: Building a knowledge base. Usability experts worked to identify key terms used by consumers and assessed the likelihood of terms used in a query for assistance conditioned on different problems.

Given such a set of terms identified at run time, we can infer the likelihood of user problems and thus appropriate help topics via Bayesian updating. At run time, we collect recognized terms from a query, reduce them to root form and input the findings for inference. We then compute the probability of all user problems (as represented by appropriate help topics), given all of the terms present $(t^+)$ and absent $(t^-)$ in a query, denoted $p(g|t^+t^-,\xi)$. Given the independence assumptions we have made, the posterior probability of each topic is given by

$$p(g_i|t^+t^-,\xi) = \\ \alpha\ p(g_i|\xi) \prod_j p(t_j^+|g_i,\xi) \prod_k 1 - p(t_k^-|g_i,\xi) \quad (1)$$

where $p(g_i|\xi)$ is the prior probability of each user informational goal $g_i$ and $\alpha$ is a renormalizing constant equal to the probability of seeing the set of terms in the phrase. We do not need to compute $\alpha$ if we wish only to generate a ranking.

## 3 Leak probabilities and assessment

Methods for limiting knowledge acquisition effort are paramount in attempts to handcraft a knowledge base for Bayesian IR. To ease the assessment load, we developed a means for focusing probability assessment on relevant terms-user problem relationships. In fleshing out the term and problem distinctions in the knowledge base, experts were asked to identify terms associated with each problem that were "relevant" or made more likely by the presence of that problem. As indicated in Figure 4, these positive influences were indicated by establishing a link between each user problem

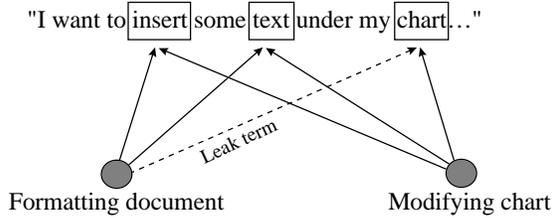

Figure 5: At run time, terms in the knowledge base are spotted. Leak terms are used for the probability of words in the knowledge base being seen, conditioned on help topics that are not directly linked to the term.

and the term positively influenced by the presence of the user problem.

During probability assessment, we explicitly assessed only the conditional probabilities of terms $t$ for user goals $g$ linked to the terms. Rather than assess conditional probabilities for terms not explicitly linked to topics, we assume a small default *leak* probability, $p(t|g,\xi) = \epsilon$, for the likelihood of seeing a term given an unlinked topic. The leak probability captures the notion that user problems may cause terms even if they are not believed to be relevant at the time the probabilistic IR model is constructed.

As captured by Figure 5, at run time, we consider whether or not terms are present in the query, and, for each user problem, whether the problem is explicitly linked to a term or not. We consider four possible outcomes for each help topic and term in the knowledge base:

1. A user problem has links to a term that is not in the query
2. A user problem has links to a term that is in the query
3. A user problem does not have links to a term that is in a query
4. A user problem does not have links to a term that is not in the query

To compute the posterior probability for a user problem, for each term appearing in the query that is not linked to that problem, we fold in the small leak probability. Thus, when a query is analyzed, we consider each of the four conditions, and compute the probability of each user problem as,

$p(g_i|t^+t^-,\xi) =$
$\alpha\ p(g_i|\xi) \prod_j p(t_j|g_i,\xi) \prod_k [1 - p(t_k|g_i,\xi)] \epsilon^l (1-\epsilon)^m$ (2)

where $t_j$ are terms seen in the query and linked to user informational goal $g_i$, $t_k$ are terms linked to user goal $g_i$ but not seen in the query, and where there are $l$ terms seen in the query but not linked to the problem $g_i$, and $m$ terms that are not linked to the problem $g_i$ and not seen in the query.

## 4 Prototyping and refinement

We worked with the Microsoft Office division to build an initial prototype of the Bayesian term-spotting methodology. The prototype was based on a knowledge base composed of approximately 600 terms to reason about the likelihood of about 40 user problems, abstracted to cover a large class of problems in the Excel domain. Informal validation of the performance of the system demonstrated that the system performed significantly better than keyword-based help systems, and a commitment was made to begin to scale the system to realistic databases covering thousands of help topics. Work on the scaling the system to such large databases and continued testing led to further improvements.

Several refinements to the basic Bayesian approach were introduced during the period of testing. Specifically, we developed additional abstractions of sets of terms that have related meaning and added new structure based in language usage in the user assistance realm.

### 4.1 Additional abstraction of terms

As part of scaling up the size of the Bayesian IR knowledge bases, we found opportunity for providing additional abstraction of terms to minimize the number of assessments and links. We developed the notion of a *metanym* to refer to sets of words that are probabilistically influenced by the existence of user problems in a similar manner. Metanyms include phrases that point at the same basic observation or concept, more closely related synonyms, hypernyms, and hyponyms. As an example, a user may employ the terms "delete," "erase," "remove," "kill," "lose," and "get rid of" in an equivalent pattern of usage. To reduce the number of links and assessments, we can define a metanym referred to as "deletion" and define it with terms that refer to the concept of deleting. Seeing any term contained in the metanym activates the metanym. The the probabilities of metanym conditioned on user problem is stored and used in the Bayesian information retrieval.

## 4.2 Modeling language about existing and desired states

In the coarse of building and testing early prototypes of the Bayesian term-spotting methodology, we identified several opportunities for refining the approach centering on a consideration of patterns in natural queries. In particular, specific cases where inappropriate topics percolated to the top of a ranked list led to valuable insights about how we might integrate additional knowledge about the relationship of user goals and language used in queries.

We found that typical queries for assistance with software frequently contain noun phrases that refer to either (1) the current state of affairs and objects that exist now, or to (2) desired states of affairs achievable via the transformation of the current state of affairs with some software functionality or through through the creation of new objects. As an example, the probabilistic relationships between "chart" and sets of user problems depend on whether "chart" is used in the phrase "this chart" versus when the word is used in the phrase "a chart." Thus, the conditional probabilities of terms given user problems may depend on whether the terms are being used to refer to *existing* or *desired* objects. Articles ("*a* chart," "*this* chart"), prepositions (e.g., "*under* this chart"), and adjectives adjacent to noun phrases appeared to be rich sources of evidence about the form of word.

We found that our intuitions about the importance of existing versus desired objects in understanding the goal of queries coincided with the notion of *definiteness* studied by linguists. Indo-European languages as well as many non-Indo-European languages, including Japanese, make use of special words and structure to communicate *existing* versus *desired* objects and states of affairs. For example in English, *definite* articles, including "the" and "this," adjacent to noun phrases typically signals that an object exists, while *indefinite* articles, such as "a" and "some," implies nonexistence of the object described in the adjacent noun phrase.

Given the potential importance of discriminating existing versus desired objects or states in interpreting queries for assistance with software applications, we developed a Bayesian approach to modeling terms used in the definite versus the indefinite sense. The approach is based on the observation that the type and number of functional words such as articles, conjunctions, prepositions, and possessives provide evidence about the probability that objects are being referred to in the definite form. For example, the use of possessives is strong evidence that the noun referred to by a function word exists. Consider the phrase, *"I'd like to change the colors of text under my chart."* The preposition, "under," and the possessive, "my," tells us that the chart is likely to exist. On the other hand, consider the phrase, *"How can I create a chart?"* The article "a" is an indefinite article, as it suggests that "chart" is a desired, but as of yet, nonexisting object.

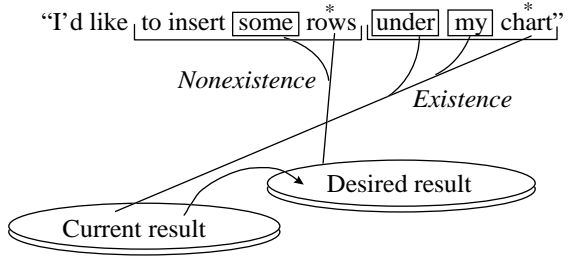

Figure 6: A Bayesian approach to considering indefiniteness in queries for assistance. We identify clauses and compute the probability of indefinite usage of terms based on adjacent function words.

We integrated an analysis of the use of the definite versus the indefinite form of a term with the basic Bayesian information retrieval methods we have described. In the approach, an expert constructing the knowledge base has the option of noting that the use of a term or metanym can be split into the indefinite and definite uses, and to indicate distinct links between user goals and the different uses of the terms. Then, conditional probabilities are assessed for the likelihood of a term being used in a query given the indefinite usage, $p(t^+|I, g_i, \xi)$, and for the definite usage of the term, $p(t^+|D, g_i, \xi)$.

Figure 6 captures the run-time analysis of the definite versus indefinite usage. When a query is analyzed, function words such as articles, possessives, and prepositions are noted and used to identify noun clauses. Then the function words adjacent to and modifying noun clauses are used to compute the probability that the noun in the adjacent clause is being used in the indefinite versus the definite sense.

To model the likelihood of indefinite usage, we consider the probability, $p(I|F_1 \ldots F_n, g, \xi)$, that the terms in a clause are being used in the indefinite form, given adjacent function words $F_1 \ldots F_n$. We can simplify the additional assessment task by assuming that $I$ is conditionally independent of the goals and construct functions that estimate the probability of indefinite use given the set of function words adjacent to the clause. If we make an additional assumption that function words are independent of other function words given $I$, we can compute $p(I|F_1 \ldots F_n, \xi)$ from an assessment of the likelihood of a function word given the presence of a noun of indefinite form, $p(F|I, \xi)$, and a

prior probability of the indefinite form, $p(I|\xi)$.

Given a knowledge base extended with assessments of the likelihoods for indefinite and definite forms of terms conditioned on goals, we use the estimated $p(I|F_1..F_n, \xi)$ to compute $p(t^+|g_i, \xi)$,

$$p(t^+|g_i,\xi) = p(t^+|I,g_i,\xi)\ p(I|F_1\ldots F_n,\xi)$$
$$+ \ p(t^+|\neg I,g_i,\xi)\ p(\neg I|F_1\ldots F_n,\xi) \qquad (3)$$

If there are no function words yielding information about the usage in the adjacent noun clause, we simply use the prior probability of the indefinite form.

We explored an extension to this method which assumed that all words are either neutral, definite, indefinite, or unknown and developed weighting approaches that appropriately adjust $p(t^+|g_i,\xi)$ for these methods. With this more detailed approach, we assumed that, if a link is not labeled, we should compute the probabilities of definite, indefinite, and neutral as reflecting the ratio of labeled links in the rest of the knowledge base.

Assessing and implementing the existence versus desired usages in the prototype led to improvements in the performance of the system. Most noticeable was the appropriate lowering of probabilities assigned to topics related to the creation of new objects when queries were issues about modifications of existing objects.

### 4.3 Disambiguating noun and verb usages

We also sought to enhance the system by adding knowledge of additional structure in language. In English, many words can be used as nouns or verbs depending on the structure of the phrase. As an example, consider the word "print" appearing in the phrase *"How do I print this?"* (verb form) or the phrase *"How can I make this print darker?"* (noun). The property describing the dual use of these words is called *zero derivation*.

We can enhance the accuracy of the Bayesian term-spotting analysis by assessing separate probabilities and links for the noun form of words and the verb form of words. That is, for a subset of specially marked zero-derivation terms, we link and assess conditional probabilities for the noun form, $p(t^+|N,g_i,\xi)$ and the verb form, $p(t^+|\neg N,g_i,\xi)$, of usage. A set of function words that appear after and before the zero-derivation terms give us deterministic knowledge on whether the term is being used as a noun or verb. At run-time, we employ templates that detect whether zero-derivation terms are being used in the noun form or verb form and return information about the use of the word. The appropriate conditional probabilities is then passed to the base probabilistic updating described above.

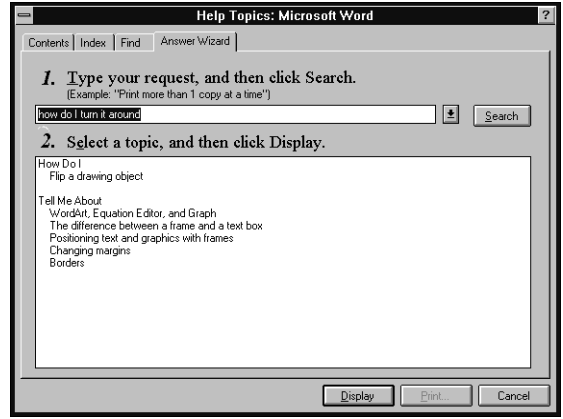

Figure 7: A derivative of the Bayesian term-spotting approach was scaled up to handle thousands of topics and served as the primary user assistance system in Microsoft Office, named *Answer Wizard*.

On rare occasions where we are uncertain of the usage, we can gather evidence about the probability that the word is being used in noun or verb form and compute the conditional probabilities of the term given goals, similar to the computation of probabilities in the context of uncertainty about the existing versus nonexisting usage. We expand the independence assumptions to assume that $g$ is independent of $N$ and $I$ given $F$, compute the probability of a term being used in the noun form, and finally compute $p(t^+|g_i,\xi)$.

## 5 Scaling the approach to the real world

Given the solid performance of the prototype, a decision was made by the Microsoft Office product division to collaborate with our team on scaling the approach up to handle thousands of topics and for the creation of distinct knowledge bases for foreign languages where Microsoft Office has a large user base. Knowledge bases were created for the Microsoft Office Suite, including Access, Excel, Powerpoint, Word, Outlook, as well as for Microsoft Project application, employing a derivative updating scheme.

As part of the effort of scaling up the approach, an assessment and testing environment was created for building knowledge bases and team of *usability experts* was assembled for enumerating distinctions and for assessing conditional probabilities. This group of people includes psychologists and other specialists in human-computer interaction.

During the early phases of assembling a team and constructing knowledge bases, numerous questions arose on the process of assessing conditional probabilities. These questions led to the development of guidelines and prototypical examples to assist with transmitting a unified vision of the nature of the probability of terms conditioned on user goals.

The implementation team found that it was helpful to allow experts to assess conditional probabilities on a scale of 1 to 13, and to later remap the assessed numbers to probabilities. The mapping between the numbers and probabilities, and associated guidelines for experts, was defined by assuming equal likelihood ratios among adjacent buckets. Definitions were provided for each point on the scale with examples and natural-language definitions that characterized the likelihood of terms, given the presence of a user goal. Guidelines were created in terms of typical sentences and parts of speech used to refer to specific goals. Such definitions assisted with explaining the task to assessors as well as for normalizing the probabilities assessed by different groups of people.

The assessment task posed a significant challenge, including formulating a distinct set of topics representing distinct user goals, identifying terms and phrases employed by users to describe their goals, to explicitly linking terms to multiple topics, and to assessing the conditional probabilities. The Bayesian IR model for the Microsoft Word application is representative for the Office applications. In Office '95, the Bayesian model for Microsoft Word included over 1,000 topics, 5,000 terms, and 145,000 dependencies.

Each full-time usability expert completed on average about 40 topics per week. After creating an English version, the models were translated into twelve languages, including German, French, Spanish, Italian, Swedish, Japanese, Brazilian Portuguese, Dutch, Danish, Norwegian, Korean, and Traditional Chinese. Localization of the English knowledge base to the other languages was found to take approximately one month for a full-time person.

To test the knowledge bases as well as to monitor for problems as real-world implementations were constructed and integrated into products, a large, covering set of "smoke test" queries was created for each software application. These queries were gleaned by usability experts from online forums, email, and studies with human subjects at Microsoft's usability labs. The performance requirement was to have a good answer for queries appear in the top five of the returned list for no less than 99 per cent of the queries.

Studies were also performed with users. In the studies, users seeking assistance were allowed to rephrase their queries up to three times for each informational goal. The studies showed that users would receive a good answer to their question within the top five recommended topics in approximately 75 percent of their queries. A large fraction of failures was attributed to user's inputting single, vague words instead of describing their goal more naturally, in a manner they might request assistance from a colleague.

# 6 Conclusion

We have described a Bayesian term-spotting methodology that allows users of computer software to request assistance by composing natural free-text queries. We presented a basic set of assumptions and a Bayesian updating method, and then described how we extended the initial approach by considering additional structure in queries.

The Bayesian framework we provided allowed for the construction and assessment of a large probabilistic information retrieval system by a team of usability experts. The costs of the intensive assessment effort required to build a large handcrafted knowledge base typically may overwhelm the value of enhanced information retrieval. However, for the case of enhancing a product used by many millions of people, the detailed assessment can be worth the cost of manual construction of a database. We are now exploring means for automating the construction of Bayesian term-spotting knowledge bases.

## Acknowledgments

We are indebted to Erich Finkelstein and Sam Hobson for alerting us to the need for enhancing access to assistance in software applications. Erich Finkelstein carried out the initial prototyping work in collaboration with researchers in our group. Sam Hobson led Microsoft Office decision making on making a commitment to implement the Bayesian term-spotting approach in Office. Adrian Klein managed the overall details of the real-world scaling up of this technology into the Answer Wizard in Microsoft Office. Kristin Dukay and Eric Hawley led the team of usability experts who built the Answer Wizard knowledge bases. We thank Jack Breese, Karen Jensen, and Greg Shaw for their feedback and comments on the methods.